\documentclass[preprint,preprintnumbers,amsmath,amssymb]{revtex4-2}
\usepackage{graphicx}
\usepackage{graphics}
\usepackage{chngcntr}
\usepackage{bm}
\usepackage{epsfig}
\begin{document}

\title{Purely leptonic decays of heavy-flavored charged mesons}

	\author{Kalpalata Dash$^{1}$\footnote{email address:kalpalatadash982@gmail.com}, P. C. Dash$^{1}$, R. N. Panda$^{1}$, Susmita Kar$^{2}$, N. Barik$^{3}$}
\address{$^{1}$ Department of Physics, Siksha $'O'$ Anusandhan Deemed to be University, Bhubaneswar-751030, India. \\$^{2}$ Department of Physics, Maharaja Sriram Chandra Bhanja Deo University,\\ Baripada-757003, India.\\ 
	$^{3}$ Department of Physics, Utkal University, Bhubaneswar-751004, India.}


\begin{abstract}
We study the purely leptonic decays of heavy-flavored charged pseudoscalar (P)  and vector (V) mesons ($D_{(s)}^{(*)+}$, $B_{(c)}^{(*)+}$) in the relativistic independent quark (RIQ) model based on an average flavor-independent confining potential in equally mixed scalar-vector harmonic form. We first compute the mass spectra of the ground-state-mesons and fix the model parameters necessary for the present analysis. Using the meson wave functions derivable in the RIQ model, and model parameters so fixed from hadron spectroscopy, we predict the decay constants: $f_{P(V)}$, ratios of decay constants: $f_{V}/f_{P}$, $f_{P_1}/f_{P_2}$, $f_{V_1}/f_{V_2}$, and the branching fractions (BFs): ${\cal B}(P(V)\to l^+\nu_l)$, $l=e, \mu, \tau$, which agree with the available experimental data and other Standard Model (SM) predictions. For the unmeasured decay constants especially in the purely leptonic decays of the charged vector mesons, our predictions could be tested in the upcoming Belle-II, SCTF, CEPC, FCC-ee and LHCb experiments in near future. 
\end{abstract}

\maketitle

\section{Introduction}	
In the last few decades, the heavy-flavored mesonic systems have attracted a great deal of attention as they provide important information on determination of fundamental parameters of the Standard Model (SM). In general, the purely leptonic charged pseudoscalar (P) and vector (V) meson decays with final lepton-neutrino pair, or lepton-lepton pair are considered as rare decays, which have simpler physics than hadronic decays. These decay rates are expressed in terms of the weak decay constants $f_{P(V)}$, which bear considerable theoretical and phenomenological importance as they govern the strength of leptonic and nonleptonic meson decays, determine the Cabbibo-Kobayashi-Maskawa (CKM) matrix elements and help in the description of the neutral $D-\bar{D}$ and $B-\bar{B}$ mixing process. The precise determination of these decay constants also helps us to test the unitarity of the quark mixing matrix and study CP violation in the SM \cite{blanke2017introduction}. 
\par The experimental measurements of decay constants for purely leptonic decays of heavy-flavored charged mesons has so far made limited progress. While the decay constants: $f_{D^+}$ \cite{artuso2005improved, eisenstein2008precision, artuso2007measurement, alexander2009measurement, ablikim2014precision, ablikim2019observation}, $f_{D_s^+}$ \cite{artuso2007measurement, alexander2009measurement, ecklund2008measurement, onyisi2009improved, zupanc2013measurements, del2010measurement, ablikim2016measurement, ablikim2021measurement, ablikim20measurement, ablikim21measurement}, $f_{B_u^+}$ \cite{hara2013evidence, kronenbitter2015measurement, lees2013evidence, aubert2010search} in the pseudoscalar sectors have so far been measured by CLEO-c, Belle, BaBar, BESIII Collaborations, $f_{B_c^+}$ is yet to be measured. In the vector meson sector only $f_{D_s^{*+}}$ have been measured recently by BESIII  Collaborations \cite{ablikim2023first} whereas $f_{D^{*+}}$ and $f_{B_{(C)}^{*+}}$ have not been measured so far. The BFs for $D_{(s)}^+\to \mu^+\nu_\mu, \tau^+\nu_\tau$ and $B_u^+\to \tau^+\nu_\tau$ have been precisely measured by  CLEO, Belle, BaBar and BESIII. For $D_{(s)}^+\to e^+\nu_e$ and  $B_u^+\to e^+\nu_e$, $\mu^+\nu_\mu$, only the upper bound of BFs are available now. This is because the BFs being proportional to $m_l^2$, these decay processes suffer from strong helicity suppression. In the vector meson sector, only the BFs of $D_s^{*+}\to e^+\nu_e$ has been measured recently by BESIII collaboration \cite{ablikim2023first}. From the currently available data statistics, it is expected that BFs for $D^{*+}\to e^+\nu_e$, $D_{(s)}^{*+}\to \mu^+\nu_\mu$, $\tau^+\nu_\tau$ could be carefully investigated at Belle-II, SCTF or STCF, CEPC, FCC-ee, LHCb future experiments. 
\par The experimental information about $b$ - flavored mesons are indeed scarce. The possibility of experimental investigation on purely leptonic decays of $B_u, B_u^*, B_c, B_c^*$ mesons discussed in \cite{yang2021purely} can be summarized as follows. Considering about $10^{13}$ Z bosons at FCC-ee \cite{abada2019fcc} and BF ${\cal B} (Z\to b\bar{b}) = 12.03\pm0.21\%$ \cite{particle2022review}, and assuming the fragmentation fraction $f (b\to B_u^*) \sim 20\%$ \cite{abdallah2003measurement}, more than $4\times 10^{11}$ $B_u^*$ events are expected which can hopefully search for the $B_u^{*+}\to l^+\nu_l$ decays. In addition, the $b$ - quark production cross section of about $\sigma(pp\to b\bar{b}X) \simeq 495$ $\mu$b at the centre-of-mass energy $\sqrt{s} = 13$ TeV is  found at LHCb \cite{aaij2015measurement, aaij2017measurement}. That is expected to yield more than $5\times 10^{13}$ $ B_u^{*}$ events with a dataset of 300 f$b^{-1}$ at LHCb and fragmentation fraction $f (b\to B_u^*) \sim 20\%$. This indicates that the $B_u^{*+} \to e^+\nu_e, \mu^+\nu_\mu, \tau^+\nu_\tau$ could be investigated at FCC-ee and LHCb future experiments. For experimental investigation of $B_c^{*+}\to l^+\nu_l$ decays, there should be at least more than $10^7$ $B_c^*$ events available. As of now, it is expected that more than $10^{12}$ Z bosons can be available at the future $e^+e^-$ colliders of CEPC \cite{dong2018derived} and FCC-ee \cite{abada2019fcc}. With BF ${\cal B} (Z\to b\bar{b}) = 12.03\pm0.21\%$ \cite{particle2022review} and fragmentation fraction $f (b\to B_c^*) \sim 6\times 10^{-4}$ \cite{boroun2016laplace, yang2019relativistic, zheng2019qcd}, there will more than $10^8$ $B_c^*$ events to search for $B_c^{*+}\to e^+\nu_e, \mu^+\nu_\mu, \tau^+\nu_\tau$ decays. In addition, the $B_c^*$ production cross sections at LHC are estimated to be about 100 nb for pp collisions at $\sqrt{s}$ = 13 TeV \cite{chen2018b} yielding more than $3\times 10^{10}$ $ B^*_c$ events corresponding to a dataset of 300 f$b^{-1}$ at LHCb. Hence, the $B_c^{*+}\to e^+\nu_e, \mu^+\nu_\mu, \tau^+\nu_\tau$ decays are expected to be carefully measured at LHCb experiments in the future.
\par An important issue in flavor physics in recent years is to test of lepton flavor universality by calculating the ratios of BFs: ${\cal{R}}_\mu^\tau$. For purely leptonic decays of charged mesons (PLDCMs): $D_{(s)}^+\to \tau^+\nu_\tau$ to $D_{(s)}^+\to \mu^+\nu_\mu$, the available experimental data \cite{particle2022review}: $({\cal{R}}_\mu^\tau)^D=\frac{{\cal{B}}(D^+\to \tau^+\nu_\tau)}{{\cal{B}}(D^+\to \mu^+\nu_\mu)}=3.21\pm 0.73$,
$({\cal{R}}_\mu^\tau)^{D_s}=\frac{{\cal{B}}(D_s^+\to \tau^+\nu_\tau)}{{\cal{B}}(D_s^+\to \mu^+\nu_\mu)}=9.82\pm 0.40$ are consistent with SM expectations. Similar observables; $(R_\mu^\tau)^{D_{(s)}^*}$, $(R_\mu^\tau)^{B_{(c)}}$ and $(R_\mu^\tau)^{B_{(c)}^*}$ have not yet to been measured.
\par The theoretical description of the PLDCMs requires a non-perturbative approach as the interactions at short distance are mediated by strong force. Within the SM, the PLDCMs are typically induced by the tree-level exchange of the gauge boson W, which subsequently decays to a charged lepton and a lepton-neutrino. The decay amplitudes represented by the decay constants $f_{P(V)}$ of the PLDCMs are evaluated from the quark and antiquark wave functions at the origin, which can not be computed from the first principle. Therefore alternate routes based on various theoretical and phenomenological approaches, have predicted these decays. Some of them are based on the nonrelativistic quark model (NRQM) \cite{ albertus2005study, yazarloo2016study, yazarloo2017mass, abu2018heavy, ebert2002decay, ebert2003properties, ebert2006relativistic, hassanabadi2016study}, relativistic quark model (RQM) \cite{ebert2002decay, ebert2003properties, ebert2006relativistic, hwang1997decay, soni2017decay, soni2019erratum, sun2017decay, sun2019wave}, Bathe-Salpeter (BS) \cite{wang2006decay, wang2004decay, cvetivc2004decay} formalism, light front quark model (LFQM) \cite{choi2007dis, choi2007, choi2015variational, chang2018decay, chang2018self, verma2012decay, hwang2010s, geng2016some, dhiman2018, dhiman2019twist, cheng2017branching}, light cone quark model (LCQM) \cite{ hwang2010analyses, dhiman2018decay}, light front holographic QCD (LFHQCD) \cite{dosch2017supersymmetry, chang2017application}, QCD sum rule (SR) \cite{gelhausen2013decay, wang2015analysis, lucha2014decay, narison2015improved, narison2016decay, narison2013fresh, baker2014b, lucha2015accurate}, and lattice QCD (LQCD) \cite{becirevic1999nonperturbatively, bowler2001decay, bevcirevic2012d, lubicz2017masses, chen2021charmed, donald2014prediction, na2012b, davies2010update, mcneile2012high, FermilabLattice:2011njy, davies2010update, dowdall2013b, dingfelder2016leptonic, zuo2024prospects, aoki2022flag} approaches. The predictions of the decay constants $f_{P(V)}$ and BFs in different theoretical and phenomenological approaches have been obtained in wide ranges. 
\par In view of the potential prospects of increased data statistics in high luminosity experiments at Belle-II, SCTF, CEPC, FCC-ee and LHCb, which are likely to yield careful measurement of the yet unmeasured decay constants and BFs, several theoretical attempts, as cited in the literature, have been taken in studying the PLDCMs. This has inspired us to undertake the present study of the PLDCMs ($D_{(s)}^+$, $B_{(c)}^+$ and $D_{(s)}^{*+}$, $B_{(c)}^{*+}$) in the framework of our relativistic independent quark (RIQ) model. Wide-ranging properties of hadrons have been described in the earlier application of the RIQ model, which includes their static properties \cite{barik1986electromagnetic, priyadarsini2016electromagnetic,  barik1998elastic, barik1987and, barik1986mass} namely the electromagnetic form factors, charge radii of pion and kaon, and pion decay constant and their decay properties \cite{ barik1995radiative,barik1998exclusive, barik1993weak, barik2008radiative, 2008radiative, barik2009radiative, patnaik2017magnetic, patnaik2018electromagnetic, barik2009semileptonic, patnaik2020semileptonic, nayak2021lepton, nayak2022exclusive, 2022exclusive, dash2023nonleptonic} such as the weak radiative, rare radiative, weak leptonic, radiative leptonic, electromagnetic, semileptonic and nonleptonic decays of $c$ - and $b$ - flavored mesons. Here, we would like to predict $f_{P(V)}$, BFs and $(R_\mu^\tau)$s and compare our results with the available experimental data and other theoretical model predictions. For the yet unmeasured decay modes, our predictions may be useful for experimental testing in the future.
\par In the present study, we consider that (1) the quark-antiquark pair inside the decaying meson-bound-state annihilate to massive W- boson, which subsequently decays to a lepton pair ($l\bar{\nu_l}$). Although, in principle, the decay can take place at any arbitrary momentum $\vec {k}$ of the parent meson, for the sake of simplicity, we consider the decay in its rest frame. We also believe that (2) there exists a strong correlation between the quark and antiquark momenta so as to have their total momentum identically zero in the decaying meson rest frame. Here, of course, an obvious difficulty arises in the context of the energy conservation at the hadron- boson vertex, since the sum total of kinetic energy of annihilating quark-antiquark pair is not equal to the rest mass-energy of the decaying meson. In the absence of any rigorous field-theoretic treatment of the quark-antiquark annihilation inside the meson, such difficulty arises as a common feature in all phenomenological model descriptions based on leading order calculation. This leads us to believe that the differential amount of energy is somehow made available to the virtual W- boson when quark-antiquark annihilation occurs with disappearance of the meson-bound state.
\par The paper is organised as follows. In Sec. II, we present the theoretical framework that includes the description of (A) the invariant transition matrix element ${\cal M}_{fi}$ leading to the general expression of the decay widths $\Gamma(P(V)\to l^+\nu_l)$, (B) a brief account of the RIQ model conventions and the quark orbitals, (C) the meson-bound-state and extraction of decay constants $f_{P(V)}$ in terms of model quantities. Sec. III is devoted to the numerical results and discussion. Finally, Sec. IV encompasses our summary and conclusion.
\section{Theoretical framework}
\subsection{Invariant transition matrix amplitude and decay width: $\Gamma(P(V)\to l^+\nu_l)$}
The leptons being free from the strong interaction, the effective Hamiltonian \cite{yang2021purely, ball2006higher, ball2007twist} for PLDCMs could be written as the product of quark current and leptonic current in the form:
\begin{equation}
    {\cal H}_{eff}=\frac{G_F}{\sqrt{2}}V_{q_1q_2}[\bar{q_1}\gamma_\mu(1-\gamma_5)q_2][\bar l \gamma^\mu (1-\gamma_5)\nu_l]+h.c.,
\end{equation}
where the contribution of the $W$-bosons is embodied in the Fermi coupling constant $G_F$ and $V_{q_1q_2}$ is the CKM matrix element between the constituent quarks of the decaying mesons. The decay amplitude is written as
\begin{eqnarray}
{\cal M}_{fi}&=&\langle \bar l \nu_l|{\cal H}_{eff}|P(V)\rangle \nonumber \\
&=&\frac{G_F}{\sqrt{2}}V_{q_1q_2}\langle \bar l \nu_l|\bar l \gamma^\mu (1-\gamma_5)\nu_l|0\rangle\langle 0|\bar{q_1}\gamma_\mu(1-\gamma_5)q_2|P(V)\rangle.   
\end{eqnarray}
The leptonic part of decay amplitude can be calculated reliably with perturbative theory. The hadronic matrix element (HME) interpolating the diquark current between the decaying meson $P(V)$ and the vacuum states, can be expressed in terms of a non-perturbative parameter: the decay constant $f_{P(V)}$ in the form:
\begin{eqnarray}
 \langle 0|\bar{q_1}(0)\gamma_\mu \gamma_5q_2(0)|P(\vec k)\rangle&=&0,\\
 \langle 0|\bar{q_1}(0)\gamma_\mu q_2(0)|P(\vec k)\rangle&=&if_{P} k_\mu, \\
     \langle 0|\bar{q_1}(0)\gamma_\mu q_2(0)|V(\vec k, \epsilon)\rangle&=&f_{V}m_{V}\epsilon_\mu, \\
    \langle 0|\bar{q_1}(0)\gamma_\mu \gamma_5q_2(0)|V(\vec k, \epsilon)\rangle&=&0.
\end{eqnarray}
Here $m_{P(V)}$, $\vec k$ are the mass, three momentum of decaying meson and $\epsilon_\mu$ is the polarization vector of the decaying vector meson (V). It is straightforward to find that the spacelike component of the HME for the purely leptonic decay of charged pseudoscalar mesons (PLDCPMs) and its timelike component for the purely leptonic decay of charged vector mesons (PLDCVMs) are zero. With the corresponding nonvanishing parts of HME, the invariant transition amplitude squared $|{\cal M}_{fi}|^2$ is expressed in terms of hadronic ($H_{00}^P$, $H_{ij}^V$) and leptonic tensor ($L_P^{00}$, $L_V^{ij}$). \\
\noindent For PLDCPMs,
\begin{eqnarray}
 H_{00}^P&=&{f^2_{P}} {m^2_{P}}\\
    L^{00}_P&=&8(\vec{k_l}\vec{k_\nu}+E_{k_l}E_{k_\nu}),
\end{eqnarray} 
 and for PLDCVMs,
    \begin{eqnarray}
 H_{ij}^V&=&{f^2_{V}} {m^2_{V}}\epsilon_i\epsilon_j^\dagger,\\
    L^{ij}_V&=&8(k_l^ik_\nu^j-k_{l\alpha}k_\nu^{\alpha}g^{ij}+k_l^jk_\nu^i+i\epsilon^{i \alpha j \beta}k_{l\alpha}k_{\nu\beta}).
\end{eqnarray}   
Then decay width $\Gamma$ in the decaying meson rest frame is calculated from the generic expression:
\begin{equation}
    \Gamma=\frac{1}{(2\pi)^2}\int \frac{d\vec k_l d\vec k_\nu}{2m_{P(V)}2E_{k_l}2E_{k_\nu}} \delta^{(4)}(k_l+k_\nu-{\hat{O}m_{P(V)}})\sum |{{\cal M}_{fi}|^2}.
\end{equation}
Here, we use two frames of reference to evaluate the contribution of Lorentz invariant pieces: hadronic and leptonic tensors. For the sake of simplicity, we evaluate the contribution of the leptonic tensor in the lepton rest frame and that of the hadronic tensor in the decaying meson rest frame to finally obtain the decay widths in the form:
\begin{eqnarray}
\Gamma(P\to l^+\nu_l)&=&\frac{G_F^2}{8\pi}|V_{q_1q_2}|^2 {f^2_{P}}{m_{P}}{m_l}^2\left(1-\frac{m_l^2}{m^2_{P}}\right)^2,\\
\Gamma(V\to l^+\nu_l)&=&\frac{G_F^2}{12\pi}|V_{q_1q_2}|^2 {m^3_{V}}{f^2_{V}}\left(1-\frac{m_l^2}{m^2_{V}}\right)^2\left(1+\frac{m_l^2}{2m^2_{V}}\right).
\end{eqnarray}
\subsection{The RIQ model conventions and quark orbitals}
In the RIQ model, a meson is picturized as a color-singlet assembly of a quark and an antiquark independently confined by an effective and average flavor-independent potential in the form:
$U(r)=\frac{1}{2}(1+\gamma^0)(ar^2+V_0)$, where ($a$, $V_0$) are the potential parameters. It is believed that the zeroth-order quark dynamics generated by the phenomenological confining potential $U(r)$, taken in equally mixed scalar-vector harmonic form, can provide an adequate tree-level description of the decay process. With the interaction potential $U(r)$ put into the zeroth-order quark lagrangian density, the ensuing Dirac equation admits a static solution of positive and negative energy as: 
\begin{eqnarray}
	\psi^{(+)}_{\xi}(\vec r)\;&=&\;\left(
	\begin{array}{c}
		\frac{ig_{\xi}(r)}{r} \\
		\frac{{\vec \sigma}.{\hat r}f_{\xi}(r)}{r}
	\end{array}\;\right){{\chi}_{ljm_j}}(\hat r),
	\nonumber\\
	\psi^{(-)}_{\xi}(\vec r)\;&=&\;\left(
	\begin{array}{c}
		\frac{i({\vec \sigma}.{\hat r})f_{\xi}(r)}{r}\\
		\frac{g_{\xi}(r)}{r}
	\end{array}\;\right)\ {\tilde \chi}_{ljm_j}(\hat r),
\end{eqnarray}
where $\xi=(nlj)$ represents a set of Dirac quantum numbers specifying 
the eigenmodes. $\chi_{ljm_j}(\hat r)$ and ${\tilde \chi}_{ljm_j}(\hat r)$
are the spin angular parts given by,
\begin{eqnarray}
\chi_{ljm_j}(\hat r) &=&\sum_{m_l,m_s}<lm_l\;{1\over{2}}m_s|
	jm_j>Y_l^{m_l}(\hat r)\chi^{m_s}_{\frac{1}{2}},\nonumber\\
	{\tilde \chi}_{ljm_j}(\hat r)&=&(-1)^{j+m_j-l}\chi_{lj-m_j}(\hat r).
\end{eqnarray}
With the quark binding energy parameter $E_q$ and quark mass parameter $m_q$ written in the form $E_q^{\prime}=(E_q-V_0/2)$, $m_q^{\prime}=(m_q+V_0/2)$ and $\omega_q=E_q^{\prime}+m_q^{\prime}$, one can obtain solutions to the radial equation for $g_{\xi}(r)$ and $f_{\xi}(r)$ in the form:
\begin{eqnarray}
	g_{nl}&=& N_{nl} \Big(\frac{r}{r_{nl}}\Big)^{l+1}\exp (-r^2/2r^2_{nl})
	L_{n-1}^{l+1/2}(r^2/r^2_{nl}),\nonumber\\
	f_{nl}&=&\frac{N_{nl}}{r_{nl}\omega_q} \Big(\frac{r}{r_{nl}}\Big)^{l}\exp (-r^2/2r^2_{nl})\nonumber\\
	&\times &\left[\Big(n+l-\frac{1}{2}\Big)L_{n-1}^{l-1/2}(r^2/r^2_{nl})
	+nL_n^{l-1/2}(r^2/r^2_{nl})\right ],
\end{eqnarray}
where $r_{nl}= (a\omega_{q})^{-1/4}$ is a state independent length parameter, $N_{nl}$
is an overall normalization constant given by
\begin{equation}
	N^2_{nl}=\frac{4\Gamma(n)}{\Gamma(n+l+1/2)}\frac{(\omega_{nl}/r_{nl})}
	{(3E_q^{\prime}+m_q^{\prime})},
\end{equation}
and
$L_{n-1}^{l+1/2}(r^2/r_{nl}^2)$ are associated Laguerre polynomials. The radial solutions yield an independent quark bound-state condition in the form of a cubic equation:
\begin{equation}
	\sqrt{(\omega_q/a)} (E_q^{\prime}-m_q^{\prime})=(4n+2l-1).
\end{equation}
\subsection{Meson-bound-state and the decay constant}
\noindent In the relativistic independent particle picture of this model, the relativistic constituent quark and antiquark are thought to move independently inside the meson-bound state. While describing any decay process in this model, we take a wave-packet representation of the meson-bound state with appropriate momentum distribution among the constituent quark and antiquark in their corresponding $SU(6)$-spin flavor configuration, since the decay takes place in the momentum eigenstate of the participating meson. The transition probability amplitude for the weak leptonic decay, calculated from the appropriate Feynman diagram, can be expressed as the free quark-antiquark pair annihilation integrated over an effective momentum distribution function  ${\cal G}(\vec{p}_{q_1}, \vec{p}_{q_{2}})$. In this model, we take ${\cal G}(\vec{p}_{q_1},\  \vec{p}_{q_{2}})=\sqrt{G_{q_1}(\vec{p}_{q_1}) G_{q_{2}}(\vec{p}_{q_{2}})}$ in the light of the ansatz of Margolis and Mendel \cite{margolis1983bag} in their bag model description of the meson-bound-state. Here, $G_{q_1}(\vec p_{q_1})$ and ${\tilde G}_{q_2}(\vec p_{q_2})$: the momentum probability amplitude of the constituent quark $q_1$ and antiquark $\bar {q_2}$, respectively; are obtained via the momentum space projection of the corresponding static solutions (quark orbitals) (14). For the ground state mesons ($n=1$,$l=0$), we find:
\begin{eqnarray}
	G_{q_1}(\vec p_{q_1})&=&{{i\pi {\cal N}_{q_1}}\over {2\alpha _{q_1}\omega _{q_1}}}
	\sqrt {{(E_{p_{q_1}}+m_{q_1})}\over {E_{p_{q_1}}}}(E_{p_{q_1}}+E_{q_1})
	\times\exp {\Big(-{
			{\vec {p}_{q_1}}^2\over {4\alpha_{q_1}}}\Big)},\\
   {\tilde G}_{{q_2}}(\vec p_{q_2})&=&-{{i\pi {\cal N}_{q_2}}\over {2\alpha _{q_2}\omega _{q_2}}}
	\sqrt {{(E_{p_{q_2}}+m_{q_2})}\over {E_{p_{q_2}}}}(E_{p_{q_2}}+E_{q_2})
	\times\exp {\Big(-{
			{\vec {p}}_{q_2}^2\over {4\alpha_{q_2}}}\Big)}.
\end{eqnarray}
With the effective momentum distribution function ${\cal G}(\vec{p}_{q_1},\  \vec{p}_{q_2})$, so obtained from model dynamics,the wavepacket representation of meson-bound-state at definite momentum $\vec{k}$ and spin projection $S_{D_{P(V)}}$ is taken in the form \cite{ barik1995radiative,barik1998exclusive, barik1993weak, barik2008radiative, 2008radiative, barik2009radiative, patnaik2017magnetic, patnaik2018electromagnetic, barik2009semileptonic, patnaik2020semileptonic, nayak2021lepton, nayak2022exclusive, 2022exclusive, dash2023nonleptonic}:
  \begin{eqnarray}
  |P(V)(\vec{k},S_{P(V)})\rangle&&={\hat{\Lambda}}(\vec{k},S_{P(V)})|(\vec{p}_{q_1},\lambda_{q_1});(\vec{p}_{q_2},\lambda_{q_2})\rangle\nonumber\\&&={\hat{\Lambda}}(\vec{k},S_{P(V)})\hat{b}^\dagger_{q_1}(\vec{p}_{q_1},\lambda_{q_1}) \hat{\tilde{b}}^\dagger_{q_2}(\vec{p}_{q_2},\lambda_{q_2})|0\rangle.
  \end{eqnarray}
Here, $|(\vec{p}_{q_1},\lambda_{q_1});(\vec{p}_{q_2},\lambda_{q_2})\rangle$ is the Fock-space representation of the bound quark and antiquark in their color-singlet configuration with respective momentum and spin: $(\vec{p}_{q_1},\lambda_{q_1})$ and $(\vec{p}_{q_2},\lambda_{q_2})$. $\hat{b}^\dagger_{q_1}(\vec{p}_{q_1},\lambda_{q_1})$ and $\hat{\tilde{b}}^\dagger_{q_2}(\vec{p}_{q_2},\lambda_{q_2})$ are the quark and antiquark creation operators and ${\hat{\Lambda}}(\vec{k},S_{P(V)})$ is a bag-like operator taken here in the integral form:
\begin{equation}
{\hat{\Lambda}}(\vec{k},S_{P(V)})=\frac{\sqrt{3}}{\sqrt{N_{{P(V)}}(\vec{k})}}\sum_{\lambda_{q_1},\lambda_{q_2}}\zeta_{q_1,{q_2}}^{P(V)}\int d\vec{p}_{q_1}\ d\vec{p}_{q_2} \delta^{(3)}(\vec{p}_{q_1}+\vec{p}_{q_2}-\vec{k})\ {\cal G}_{{P(V)}}(\vec{p}_{q_1}, \vec{p}_{q_2}),       
\end{equation}
where $\sqrt{3}$ is the effective color factor, $\zeta_{q_1,{q_2}}^{P(V)}$ is the $SU(6)$ spin-flavor coefficients for the meson state $|{P(V)}(\vec k, S_{P(V)})\rangle$ and $N_{{P(V)}}(\vec{k})$ is the meson state normalization, which is obtained in an integral form:
\begin{equation}
 N_{P(V)}(\vec{k})=\frac{1}{(2\pi)^3{2E_k}}\int d\vec{p}_{q_1} |{\cal G}_{{P(V)}}(\vec{p}_{q_1}, \vec p_{q_2})|^2=\frac{{\bar{N}_{P(V)}(\vec{k})}}{(2\pi)^3{2E_k}}.     
\end{equation} 
Using the wave-packet representation of the meson-bound-state (21-23), it is straightforward to calculate the decay constants $f_{P}$ and $f_{V}$ from Eq. (4) and (5), respectively in the form:
\begin{eqnarray}
f_{P}=&&\frac{2\sqrt{3}}{\sqrt{(2\pi)^3m_{P}{\bar{N}_{P}(0)}}}{\int \frac{d\vec{p}_{q_1}}{\sqrt{2E_{p_{q_1}}2E_{-{p_{q_1}}}}}} \ \  {{\cal G}_{P}(\vec{p}_{q_1}, -\vec{p}_{q_1}})\nonumber \\
&&\times{\left[\frac{|\vec p_{q_1}|^2-(E_{p_{q_1}}+m_{q_1})(E_{-{p_{q_1}}}+m_{q_2})}{ \sqrt{(E_{p_{q_1}}+m_{q_1})(E_{-{p_{q_1}}}+m_{q_2})}}\right]},  
\end{eqnarray}
and 
\begin{eqnarray}
f_{V}=&&\frac{2\sqrt{3}}{\sqrt{(2\pi)^3m_{V}{\bar{N}_{V}(0)}}}{\int \frac{d\vec{p}_{q_1}}{\sqrt{2E_{p_{q_1}}2E_{-{p_{q_1}}}}}} \ \ {{\cal G}_{V}(\vec{p}_{q_1}, -\vec{p}_{q_1}})\nonumber \\
&&\times{\left[\frac{|\vec p_{q_1}|^2+3(E_{p_{q_1}}+m_{q_1})(E_{-{p_{q_1}}}+m_{q_2})}{ 3\sqrt{(E_{p_{q_1}}+m_{q_1})(E_{-{p_{q_1}}}+m_{q_2})}}\right]}. 
\end{eqnarray}
\section{Numerical results and Discussion}
In this section, we calculate the decay constants $f_{P(V)}$, BFs and ratio of BFs: ($R_\mu^\tau$) for purely leptonic decays of charged pseudoscalar and vector mesons: $(D^+_{(s)}, B_{(c)}^+; D^{*+}_{(s)}, \\ B_{(c)}^{*+})$. In our numerical calculation, we use the input parameters: the quark masses ($m_u=m_d$, $m_s$, $m_c$, $m_b$) and potential parameters ($V_0$, $a$). By using suitable values of the input parameters, we generate the ground state hyperfine mass splitting of the heavy flavored mesons: ($D^{*+}$, $D^+$), ($D^{*+}_s$, $D^{+}_s$), ($ B_u^{*+}$, $ B_u^{+}$), ($B_{c}^{*+}$, $B_{c}^{+}$), ($B_{s}^{*0}$, $B_{s}^{0}$), ($J/\psi$, $\eta_c$) and ($\Upsilon$, $\eta_b$) by appropriately taking into account, the centre of mass correction and one gluon exchange correction as per Ref. \cite{barik1986mass, barik1987and}. Our input parameters so fixed from hadron spectroscopy, are shown in Table I. \\ 
\begin{table}[!hbt]
	\renewcommand{\arraystretch}{0.1}
	\centering
	\setlength\tabcolsep{2pt}
	\caption{The quark masses $m_q$ (in $GeV$) and potential parameters: $V_0$ (in $GeV$) and $a$ (in $GeV^3$)} 

	\label{tab1}
	
		\begin{tabular}{cccccc}
			\hline
   \hline
   &&&&&\\
 \ \ $m_u=m_d$& \ \ \ $m_s$& \ \ \ $m_c$& \ \ \ $m_b$& \ \ \ $a$ \ \ &$\ \ V_0$ \ \ \\
&&&&&\\
 \ \ 0.26& \ \ \ 0.49& \ \ \ 1.64& \ \ \ 4.92&\ \ \ 0.023& \ \ -0.307\\
 &&&&&\\
 \hline
  \hline
\end{tabular}
	
\end{table}

\begin{table}[!hbt]
	\renewcommand{\arraystretch}{0.1}
	\centering
	\setlength\tabcolsep{10pt}
	\caption{Hyperfine splitting of the ground-state heavy flavored mesons with the quark-gluon coupling constant $\alpha_s=0.37$}
	\label{tab2}
	
		\begin{tabular}{ccccl}
			\hline
			\hline
			\ \ \ Meson \ \ \ \ \ \  &\multicolumn{2}{c}{Spin-averaged mass}  \ \ \ \ \  &  \multicolumn{2}{c}{Meson mass}\\
   &\multicolumn{2}{c}{ (MeV)} & \multicolumn{2}{c}{(MeV)}\\
   &&&&\\
			&Theory&Expt.& \ \ \ Theory \ \ \ \ 
   & \ \ Expt.\\
   &&&&\\
		\hline
			&&&&\\
		$D^{*\pm}$&&&1979.95&2010.26\\
		&1954.93&1975.07&&\\
		$D^{\pm}$&&&1889.83&1869.50\\
		&&&&\\
		$D_s^{*\pm}$&&&2090.38&2112.20\\
		&2067.33&2076.40&&\\
		$D_s^{\pm}$&&&1998.18&1969.0\\
		&&&&\\
		$B_u^{*\pm}$&&&5300.11&5324.71\\
		&5290.04&5313.35&&\\
		$B_u^{\pm}$&&&5239.04&5279.25\\
		&&&&\\
		$B_c^{*\pm}$&&&6290.90& \ \ \ -----\\
		&6288.43&-----&&\\
		$B_c^{\pm}$&&&6264.46&6274.47\\
		&&&&\\
            $B_s^{*0}$&&&5395.02&5415.80\\
		&5385.39&5403.58&&\\
		$B_s^{0}$&&&5360.59&5366.91\\
		&&&&\\
		$J/\psi$&&&3051.55&3096.90\\
		&3037.68&3068.65&&\\
		$\eta_c$&&&3012.16&2983.90\\
			&&&&\\
		$\Upsilon$&&&9447.06&9460.40\\
		&9443.46&9444.98&&\\
		$\eta_b$&&&9421.95&9398.70\\
			&&&&\\
			\hline
			\hline
			
		\end{tabular}
	
\end{table}
\begin{table}[!hbt]
 	\renewcommand{\arraystretch}{1.25}
 	\centering
 	\setlength\tabcolsep{12pt}
 	\caption{Numerical Inputs}
 	\label{tab3}
 	\begin{tabular}{cllc}	
  \hline
 		\hline Parameter& \ \ \ \ \ \ Value& Unit& Reference \\
\hline
$G_F$&$1.1663788\times 10^{-5}$&GeV$^{-2}$&\cite{particle2022review}\\	
$|V_{cd}|$&$0.22486\pm 0.00067 $ &......&\cite{particle2022review}\\	
$|V_{cs}|$&$0.97349\pm 0.00016$&......&\cite{particle2022review}\\	
$|V_{ub}|$&$0.00369\pm 0.00011 $ &......&\cite{particle2022review}\\	
$|V_{cb}|$&$0.04182^{+0.00085}_{-0.00074}$&......&\cite{particle2022review}\\
$\tau_{D^{ +}}$& $1033\pm5$&fs&\cite{particle2022review}  \\
$\tau_{D_s^{+ }}$& $504\pm4$&fs&\cite{particle2022review} \\
$\tau_{B_u^{ +}}$& $1.638\pm0.004$&ps&\cite{particle2022review}  \\
$\tau_{B_c^{+ }}$& $0.51\pm0.009$&ps&\cite{particle2022review} \\
$\Gamma^{total}_{D^{* +}}$& $83.4\pm1.8$&keV&\cite{particle2022review}  \\
\hline
 		\hline
 	\end{tabular}
 \end{table}
 \begin{figure}[!hbt]
	\centering
	\includegraphics[width=0.45\textwidth]{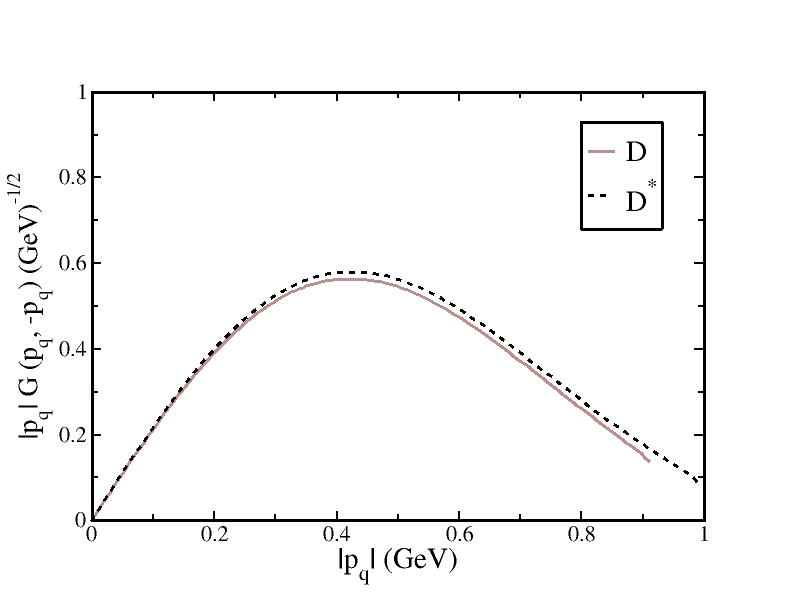}
	\includegraphics[width=0.45\textwidth]{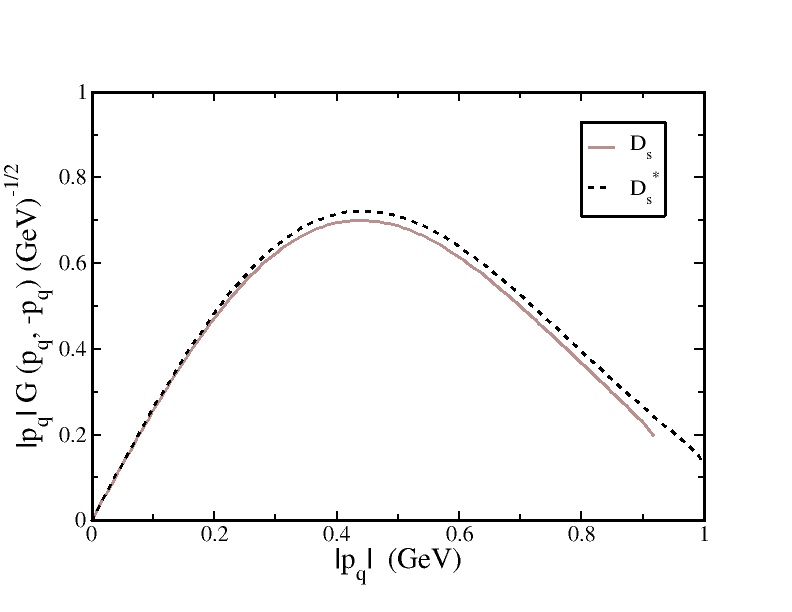}
	\includegraphics[width=0.45\textwidth]{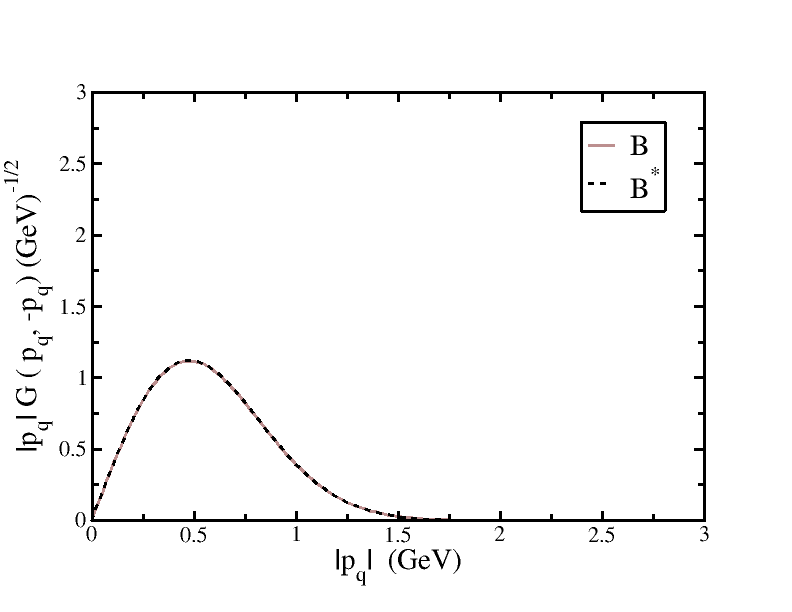}
         \includegraphics[width=0.45\textwidth]{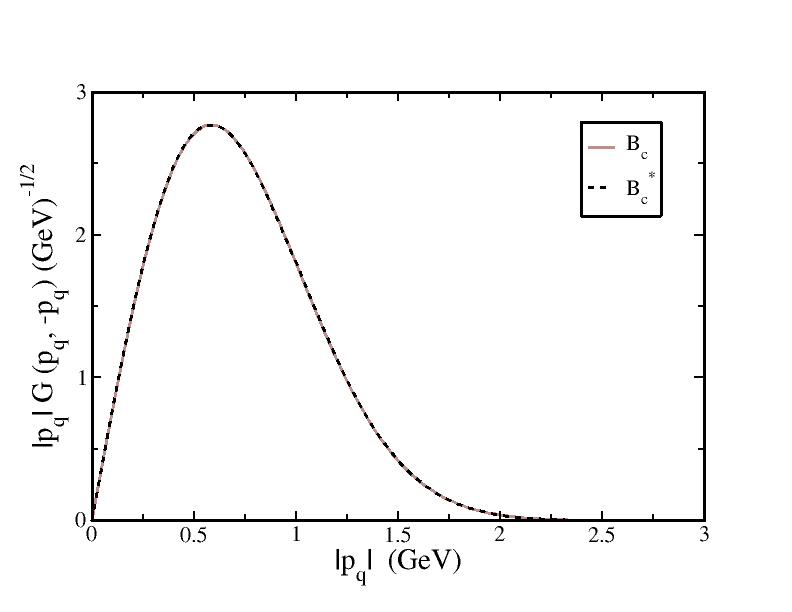}
	\caption{Radial quark momentum distribution amplitude for heavy pseudoscalar meson (solid line) and vector meson (dotted line)}
\end{figure}
\begin{figure}[!hbt]
	\centering
        \includegraphics[width=0.47\textwidth]{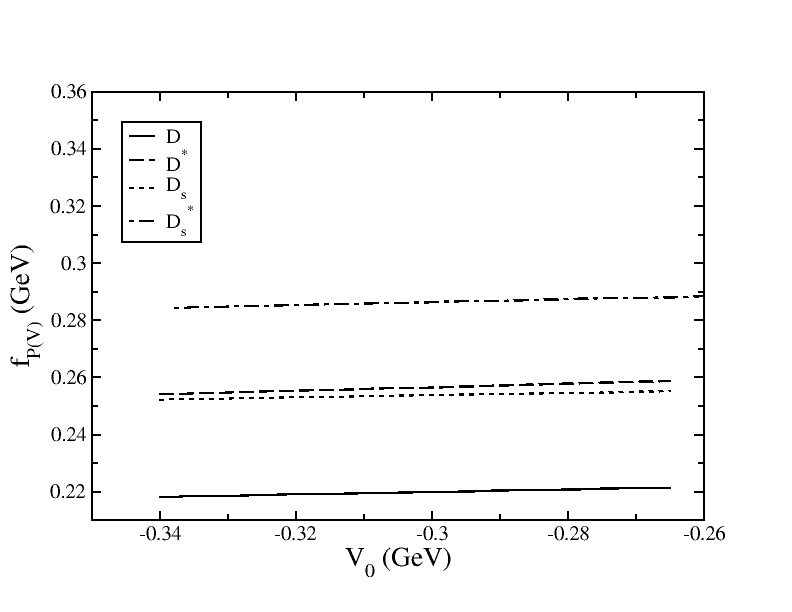}
         \includegraphics[width=0.47\textwidth]{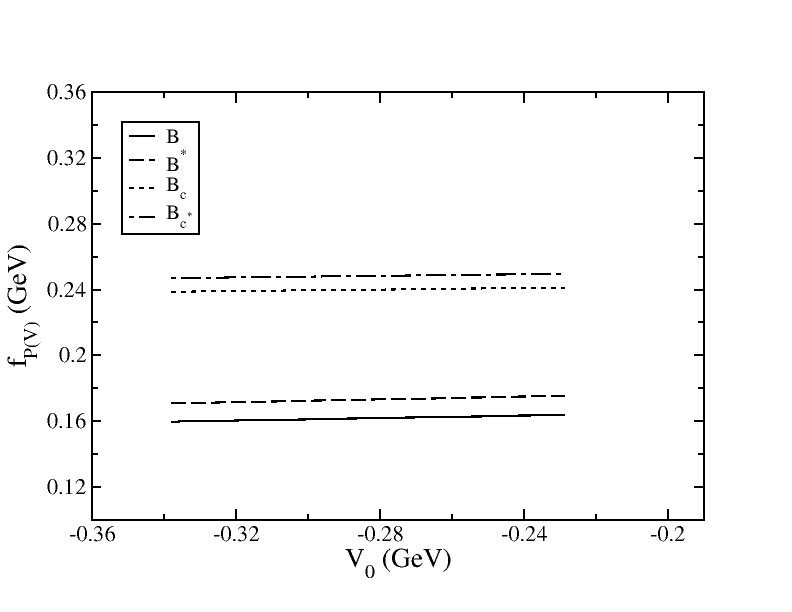}
	\includegraphics[width=0.47\textwidth]{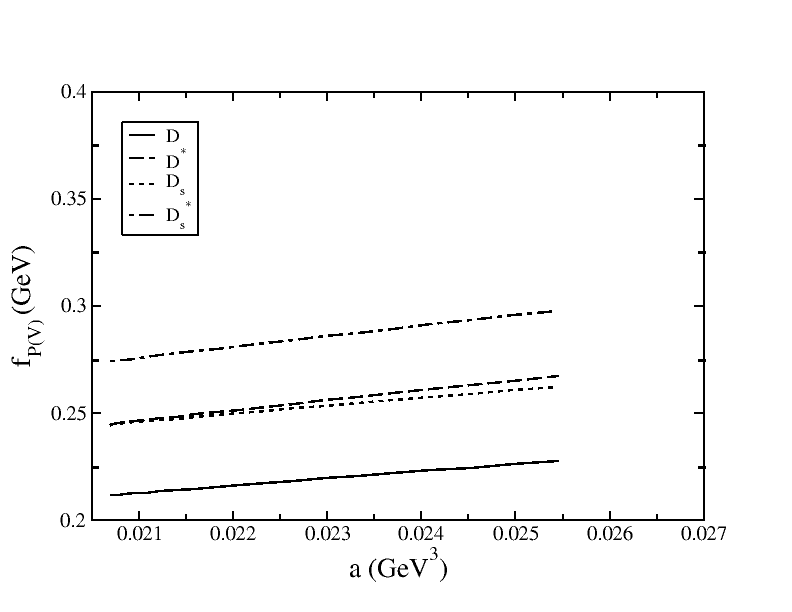}
	\includegraphics[width=0.47\textwidth]{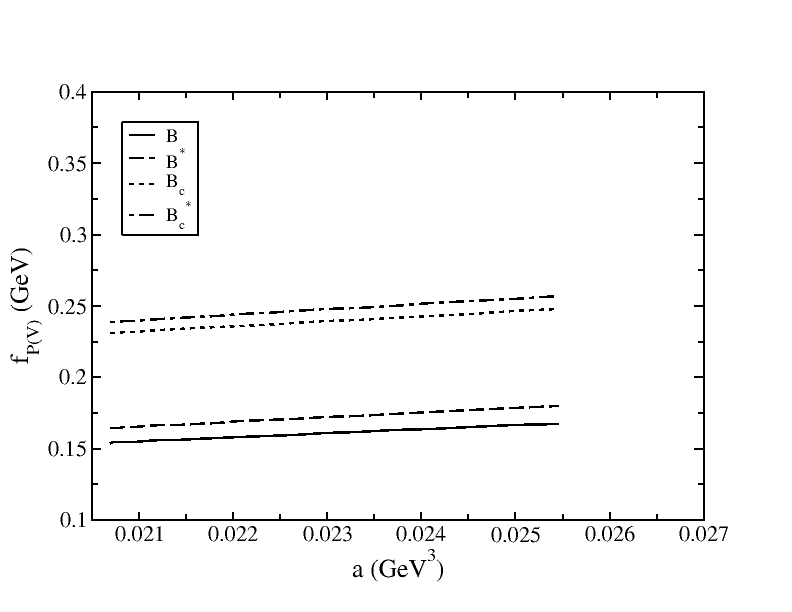}
	
	\caption{Dependence of decay constants $f_{P(V)}$ on the potential parameters ($V_0, a$)}
\end{figure}
\par The resulting ground-state-meson masses in the RIQ model are cited in Table II in good comparison with the available experimental data. The phenomenological input parameters such as the Fermi Coupling constant $G_F$, lifetime $\tau$ of pseudoscalar mesons $(D_{(s)}, B_{(c)})$ and total decay width $\Gamma^{total}_{D^{*}}$ are taken from PDG \cite{particle2022review}. For CKM parameters, we take their precise global fit values, also from PDG  \cite{particle2022review}. In the absence of the observed mass of $B_c^*$-meson, we use our model predicted value $m_{B_c^*}=6.2909$ GeV (Table II). 
\par Before calculating numerically, the physical quantities of interest: the decay constants $f_{P(V)}$ and corresponding BFs, we study the behaviour of radial quark momentum distribution amplitude $|{\vec p}_q| {\cal G}_{P(V)} ({\vec p}_q, -{\vec p}_q)$ for decaying meson-ground-state $|P(V)\rangle$, over the allowed physical range of the quark momentum $|{\vec p}_q|$. The behaviours of $|{\vec p}_q| {\cal G}_{P(V)} ({\vec p}_q, -{\vec p}_q)$ shown in Fig. 1 indicate sharper peaks for  $(B_u, B_u^*)$ and $(B_c, B_c^*)$ than those for $(D, D^*)$ and $(D_s, D_s^*)$ meson states. This is due to the large mass difference between $b$ and $c$ quarks. The quark momentum distribution amplitude for $(D_s, D_s^*)$, $(B_c, B_c^*)$ are also found to have comparatively sharper peaks than $(D, D^*)$, $(B_u, B_u^*)$, respectively. This is because of the dominant contributions of heavier quark masses $m_s$ and $m_c$ as compared to that of $m_d$ and $m_u$, respectively. As one may expect from heavy quark symmetry, the difference between the behaviour of $|{\vec p}_q| {\cal G}_{P(V)} ({\vec p}_q, -{\vec p}_q)$ of the heavier pseudoscalar $(B_u, B_c)$ and vector $(B_u^*, B_c^*)$ mesons get reduced in contrast to that of comparatively lighter pseudoscalar $(D, D_s)$ and vector $(D^*, D_s^*)$ meson in the charm sector. 
\par Using the input parameters (Table I) which include the quark masses $m_q$ and potential parameters ($V_0, a$), we calculate numerically the decay constants $f_{P(V)}$. Since the quark masses vary from one decaying meson state to other, we study the dependence of $f_P$ and $f_V$ over a chosen range of variation of potential parameters ($V_0, a$). The behaviour of the decay constants $f_{P(V)}$ over $\pm 10\%$ variation of potential parameter values are shown in Fig. 2. Our predicted decay constants and ratios of decay constants are shown in Tables (IV-VII). In order to study the sensitivity of input parameters in our model calculation, we include the systematic errors in the analysis, obtained both from the $\pm10\%$ variation of the potential parameters ($V_0, a$) for fixed quark masses ($m_q$) and $\pm10\%$ variation of the quark masses for fixed values of potential parameters. The first errors in our model calculations (Table IV-IX) come from the $\pm10\%$ variation of quark masses for fixed potential parameters and the second errors come from the $\pm10\%$ variation of potential parameters for fixed quark masses, respectively. As one can see that our predictions for PLDCVMs are more sensitive to the variation of potential parameters than those of quark masses whereas reverse is found true for PLDCPMs.
\begin{table}[!hbt]
 	\renewcommand{\arraystretch}{1.25}
 	\centering
 	\setlength\tabcolsep{5pt}
 	\caption{Predicted decay constants for ($D, D_s, D^*, D_s^*$) mesons (in MeV) compared to available experimental data and other theoretical predictions.}
 	\label{tab4}
  \resizebox{\textwidth}{!}
	{
 	\begin{tabular}{l|l|l|l|l}
 		\hline
 		\hline  Reference&$f_{D^{+}}$& $f_{D^{*+}}$&  $f_{D_s^{+}}$& \ \ $f_{D_s^{*+}}$ \\
\hline
This work&$219.58^{+10.72+8.76}_{-11.49-9.33}$&$256.09^{+7.49+12.45}_{-7.79-13.03}$&$253.50^{+13.12+9.46}_{-14.03-10.06}$&$285.97^{+9.92+12.75}_{-10.38-13.37}$\\
Expt. \cite{ablikim2023first} & {\bf ------}& {\bf ------}& {\bf ------}&$213.6^{+61.0}_{-45.8_{stat.}}{\pm 43.9}_{syst.}$\\
Expt. \cite{particle2022review} &$203.8\pm4.7\pm0.6\pm1.4$& {\bf ------}&$250.1\pm2.2\pm0.04\pm1.8$& {\bf ------}\\
LFQM Lin \cite{choi2007} &197&239&233&274\\
LFQM HO \cite{choi2007}&180&212&218&252\\
LFQM Lin \cite{choi2015variational}&208&230&231&260\\
LFQM \cite{dhiman2019twist}&$197_{-20-1.0}^{+19+0.2}$&$230_{-28+6}^{+29-5}$&$219_{-22-0.8}^{+21-0.2}$&$253_{-31+6}^{+31-6}$\\
LCQM \cite{dhiman2018}&209&260&237&291\\
LFHQCD \cite{dosch2017supersymmetry}&199&{\bf ------}&216&{\bf ------}\\
LFHQCD \cite{chang2017application}&$214.2^{+7.6}_{-7.8}$&{\bf ------}&
$253.5^{+7.6}_{-7.8}$&{\bf ------}\\
QCD SR \cite{gelhausen2013decay}&$201^{+12}_{-13}$&$242^{+20}_{-12}$&$238^{+13}_{-23}$&$293^{+19}_{-14}$\\
QCD SR \cite{wang2015analysis}&$208\pm10$&$263\pm21$&$240\pm10$&$308\pm21$\\
LQCD \cite{bowler2001decay}&$206\pm4_{-10}^{+17}$&$234\pm26$&$229\pm3_{-12}^{+23}$&$254\pm17$\\
LQCD \cite{lubicz2017masses}&$207.4\pm3.8$&$223.5\pm8.4$&$247.2\pm4.1$&$268.8\pm6.6$\\
LQCD \cite{chen2021charmed}&$213\pm5$&$234\pm6$&$249\pm7$&$274\pm7$\\
LQCD \cite{aoki2022flag}&$211.9\pm1.1$& {\bf ------}&$249\pm1.2$& {\bf ------}\\
BS \cite{wang2006decay}&$230\pm25$&$340\pm23$&$248\pm 27$&$375\pm24$\\
BS \cite{wang2004decay}&238&{\bf ------}&241&{\bf ------}\\
RQM \cite{ebert2006relativistic}&234&310&268&315\\
RQM \cite{hwang1997decay}&$271\pm14$&$327\pm13$&$309\pm15$&$362\pm15$\\
NRQM \cite{yazarloo2017mass}&368.8&353.8&394.8&382.1\\
NRQM \cite{abu2018heavy}&220&290&250&310\\
NRQM \cite{hassanabadi2016study}&228&{\bf ------}&273&{\bf ------}\\
\hline
 		\hline
 	\end{tabular}
  }
 \end{table}

 \begin{table}[!hbt]
 	\renewcommand{\arraystretch}{1.25}
 	\centering
 	\setlength\tabcolsep{5pt}
 	\caption{Ratio of decay constants for  $(D, D_s, D^*, D_s^*)$ mesons compared with the available experimental data and other theoretical predictions.}
 	\label{tab5}
  \resizebox{\textwidth}{!}
	{
 	\begin{tabular}{l|l|l|l|l}
 		\hline
 		\hline  Reference&${f_{D^{*+}}}/{f_{D^{+}}}$&${f_{D_s^{*+}}}/{f_{D_s^{+}}}$&  ${f_{D_s^{+}}}/{f_{D^{+}}}$&  ${f_{D_s^{*+}}}/{f_{D^{*+}}}$ \\
\hline
This work&$1.166^{-0.022+0.010}_{+0.027-0.010}$&$1.128^{-0.018+0.008}_{+0.023-0.008}$&$1.154^{+0.003-0.003}_{-0.004+0.003}$&$1.117^{+0.006-0.004}_{-0.007+0.005}$\\
Expt. \cite{particle2022review} & {\bf ------}& {\bf ------}& $1.228\pm 0.03\pm 0.004\pm0.009$&{\bf ------}\\
LFQM Lin \cite{choi2007} &1.21&1.18&1.18&1.15\\
LFQM HO \cite{choi2007}&1.18&1.16&1.21&1.19\\
LFQM Lin \cite{choi2015variational}&1.11&1.13&1.11&1.13\\
LFQM \cite{dhiman2019twist}&$1.17^{+0.03-0.03}_{-0.03+0.04}$&$1.16^{+0.03-0.03}_{-0.03+0.03}$&$1.11^{-0.001-0.002}_{+0.001+0.002}$&$1.10^{-0.003-0.002}_{-0.001-0.003}$\\
LCQM \cite{dhiman2018}&1.24&1.23&1.13&1.12\\
LFHQCD \cite{dosch2017supersymmetry}&{\bf ------}&{\bf ------}&1.09&{\bf ------}\\
LFHQCD \cite{chang2017application}&{\bf ------}&{\bf ------}&
$1.184^{+0.054}_{-0.052}$&{\bf ------}\\
QCD SR \cite{gelhausen2013decay}&$1.20^{+0.13}_{-0.07}$&$1.21^{+0.13}_{-0.05}$&$1.18^{+0.04}_{-0.05}$&$1.21\pm0.05$\\
QCD SR \cite{wang2015analysis}&{\bf ------}&{\bf ------}&$1.15\pm0.06$&{\bf ------}\\
LQCD \cite{bowler2001decay}&{\bf ------}&{\bf ------}&$1.11\pm0.1_{-1}^{+1}$&{\bf ------}\\
LQCD \cite{lubicz2017masses}&$1.078\pm0.036$&$1.087\pm0.020$&{\bf ------}&{\bf ------}\\
LQCD \cite{chen2021charmed}&$1.10\pm0.03$&$1.10\pm0.04$&$1.16\pm0.03$&$1.17\pm0.03$\\
BS \cite{wang2006decay}&{\bf ------}&{\bf ------}&$1.08\pm0.01$&$1.10\pm0.06$\\
BS \cite{wang2004decay}&{\bf ------}&{\bf ------}&1.01&{\bf ------}\\
RQM \cite{ebert2006relativistic}&1.32&1.18&1.15&1.02\\
RQM \cite{hwang1997decay}&$1.21\pm0.02$&$1.17\pm0.02$&$1.14\pm0.01$&{\bf ------}\\
NRQM \cite{yazarloo2017mass}&0.96&0.97&1.07&1.08\\
NRQM \cite{abu2018heavy}&1.32&1.24&1.14&1.07\\
NRQM \cite{hassanabadi2016study}&{\bf ------}&{\bf ------}&1.20&{\bf ------}\\
\hline
 		\hline
 	\end{tabular}
  }
 \end{table}
 \par Our predictions for decay constants: $f_{D^+}$, $f_{D^{*+}}$, $f_{D_s^+}$, $f_{D_s^{*+}}$ are shown in Table IV in comparison with the available experimental data and other SM predictions. It is evident that our predicted $f_{D^+}$, $f_{D_s^+}$, $f_{D_s^{*+}}$ agree with corresponding observed data within their experimental limits, and are in good comparison with the predictions obtained in the calculation based on NRQM \cite{abu2018heavy, hassanabadi2016study}, BS \cite{wang2006decay, wang2004decay}, LFQM \cite{choi2015variational, dhiman2019twist}, LFHQCD \cite{chang2017application}, QCD SR \cite{gelhausen2013decay, wang2015analysis}, and LQCD \cite{bowler2001decay, lubicz2017masses, chen2021charmed} approaches. Our predicted $f_{D^{*+}}$ is also in good comparison with the predictions of LFQM Lin \cite{choi2007}, LFQM \cite{dhiman2019twist}, LCQM \cite{dhiman2018}, QCD SR \cite{gelhausen2013decay, wang2015analysis} and the LQCD \cite{bowler2001decay, lubicz2017masses, chen2021charmed} approaches. It is interesting to predict the ratios $f_V/f_P$, $f_{P_1}/f_{P_2}$ and $f_{V_1}/f_{V_2}$, which are sensitive to the difference between vector ($V$) and pseudoscalar ($P$), pseudoscalar $(P_1)$ and pseudoscalar $(P_2)$ and vector $(V_1)$ and vector $(V_2)$ wave functions, respectively. Our predictions of ratios ($f_V/f_P$, $f_{P_1}/f_{P_2}$ and $f_{V_1}/f_{V_2}$) are given in Table V and VII.
\par In Table V, we provide our result ${f_{D^{*+}}}/{f_{D^{+}}}=1.166^{-0.022+0.010}_{+0.027-0.010}$ which is in good comparison with the results obtained from RQM \cite{hwang1997decay}, LFQM HO \cite{choi2007}, LFQM \cite{dhiman2019twist} and QCD SR \cite{gelhausen2013decay} approaches. Our prediction of ${f_{D_s^{*+}}}/{f_{D_s^{+}}}=1.128^{-0.018+0.008}_{+0.023-0.008}$ exactly matches with 1.13 of LFQM Lin \cite{choi2015variational} and also agrees well with many other SM predictions including those of RQM \cite{hwang1997decay}, LFQM \cite{choi2007, dhiman2019twist}, QCD SR \cite{gelhausen2013decay} and LQCD\cite{chen2021charmed}. Our results for the ratio ${f_{D_s^{+}}}/{f_{D^{+}}}=1.154^{+0.003-0.003}_{-0.004+0.003}$ is not only comparable to the available experimental data \cite{particle2022review} but also consistent with other SM predictions from RQM \cite{ebert2006relativistic, hwang1997decay}, LFHQCD \cite{chang2017application} and the QCD SR \cite{gelhausen2013decay, wang2015analysis}. Our result ${f_{D_s^{*+}}}/{f_{D^{*+}}}=1.117^{+0.006-0.004}_{-0.007+0.005}$ is in good agreement with the corresponding results from LFQM Lin \cite{choi2015variational} and LCQM \cite{dhiman2018}.

\begin{table}[!hbt]
 	\renewcommand{\arraystretch}{1.25}
 	\centering
 	\setlength\tabcolsep{10pt}
 	\caption{Predicted decay constants for $(B_u, B_c, B_u^*, B_c^*)$ mesons (in MeV) compared to available experimental data and other theoretical predictions.}
 	\label{tab6}
  \resizebox{\textwidth}{!}
	{
 	\begin{tabular}{l|l|l|l|l|l}
 		\hline
 		\hline  Reference&$f_{B_u^{+}}$& $f_{B_u^{*+}}$& Reference& $f_{B_c^{+}}$&  $f_{B_c^{*+}}$ \\
\hline
This work&$161.34^{+7.42+5.8}_{-7.81-6.14}$&$172.61^{+4.9+8.54}_{-5.06-8.93}$&This work&$249.50^{+10.68+9.29}_{-11.45-9.85}$&$258.66^{+9.85+10.1}_{-10.47-10.68}$\\
Expt. \cite{particle2016review} &$188\pm17\pm18$& {\bf ------}&LFQM Lin \cite{choi2015variational}&$389^{+16}_{-3}$&$391^{-5}_{+4}$ \\
LFQM Lin \cite{choi2007} &171&186&LCDA \cite{hwang2010analyses}&360&387\\
LFQM HO \cite{choi2007}&161&173&QCD SR \cite{aliev2019properties}&$270\pm30$&$300\pm30$\\
LFQM Lin \cite{choi2015variational}&181&188& QCD SR \cite{narison2020spectra}& $371\pm17$& $442\pm44$\\
LFQM \cite{dhiman2019twist}&$163_{-20+4}^{+21-4}$&$172_{+24-6}^{-23+6}$&BS \cite{wang2006decay}&{\bf ------}&$418\pm 24$\\
LCQM \cite{dhiman2018}&193&211&&&\\
LFHQCD \cite{dosch2017supersymmetry}&194&{\bf ------}&&\\
LFHQCD \cite{chang2017application}&$191.7^{+7.9}_{-6.5}$&{\bf ------}&&&\\
LQCD \cite{bowler2001decay}&$195\pm6_{-23}^{+24}$&$190\pm28$&&&\\
QCD SR \cite{gelhausen2013decay}&$207^{+17}_{-9}$&$210^{+10}_{-12}$&&&\\
QCD SR \cite{wang2015analysis}&$194\pm15$&$213\pm18$&&&\\
BS \cite{wang2006decay}&$196\pm29$&$238\pm28$&&&\\
BS \cite{wang2004decay}&193&{\bf ------}&&&\\
RQM \cite{ebert2006relativistic}&189&219&&&\\
RQM \cite{hwang1997decay}&$231\pm9$&$252\pm10$&&&\\
NRQM \cite{yazarloo2017mass}&235.9&234.7&&&\\
NRQM \cite{abu2018heavy}&147&196&&&\\
NRQM \cite{hassanabadi2016study}&149&{\bf ------}&&&\\
\hline
 		\hline
 	\end{tabular}
  }
 \end{table}
 \par In Table VI, we present our predicted decay constants of $B_u, B_c, B_u^*, B_c^*$ mesons. Our result: $f_{B_u^{+}}=161.34^{+7.42+5.8}_{-7.81-6.14}$ MeV not only agrees with the observed data \cite{particle2016review} within the experimental limits but also compares well with the prediction based on BS \cite{wang2006decay}, LFQM \cite{choi2007, dhiman2019twist}, LQCD \cite{bowler2001decay} approaches. In the absence of observed data in $B_u^{*+}$ sector, we find that our prediction $f_{B_u^{*+}}=172.61^{+4.9+8.54}_{-5.06-8.93}$ MeV compares well with those of LFQM \cite{dhiman2019twist, dhiman2019twist} and the LQCD \cite{bowler2001decay}. In the $B_c$ sector, however, the experimental measurements have been too scarce. Not much theoretical attempts have also been made in this sector to adequately address the issue. We therefore, compare our prediction on $f_{B_c^{+}}$,  $f_{B_c^{*+}}$ with a few SM predictions available in the literature. Although our result $f_{B_c^{+}}=249.50^{+10.68+9.29}_{-11.45-9.85}$ MeV is in reasonable agreement with QCD SR \cite{aliev2019properties}, our prediction $f_{B_c^{*+}}=258.66^{+9.85+10.1}_{-10.47-10.68}$ MeV appears somewhat underestimated compared to those obtained in BS \cite{wang2006decay}, LFQM Lin \cite{choi2015variational}, LCDA \cite{hwang2010analyses} and QCD SR \cite{ narison2020spectra} approaches. Note that both the valence quarks of $B_c^{(*)}$ mesons being heavy quarks, their Compton wavelengths $\sim1/m_{b,c}$ are much shorter than typical hadron size. In this scenario, we predict an approximate relation $f_{B_c}\simeq f_{B_c^*}$, as expected from spin-flavor symmetry in the heavy quark limit. 
\par As can be seen from Table VII, our predicted ratio of decay constants: $f_{B_u^{*+}}/f_{B_u^{+}}=1.069_{+0.021-0.015}^{-0.018+0.014}$ agrees well with that of LFQM Lin \cite{choi2015variational},  LFQM Lin \cite{choi2007} and LFQM \cite{dhiman2019twist} and is comparable to that of RQM \cite{hwang1997decay}, QM Lin \cite{choi2015variational}. In $B_c^*$ sector, our predicted $f_{B_c^{*+}}/f_{B_c^{+}}$ is also found in reasonable agreement with LFQM Lin \cite{choi2015variational}, QCD SR \cite{aliev2019properties, narison2020spectra}. Finally, we predict $f_{B_c^{*+}}/f_{B_u^{*+}}=1.498^{+0.014-0.015}_{-0.017+0.017}$ and $f_{B_c^{+}}/f_{B_u^{+}}=1.546^{-0.005+0.002}_{+0.007-0.002}$ which can be verified in the upcoming experimental measurements and compared with future SM predictions in this sector.
\par With the predictions of decay constants $f_P$ and $f_V$ in our model framework and other necessary phenomenological input parameters shown in Table II, it is straightforward to calculate the decay rates for PLDCPMs ($D^+_{(s)}, B^+_{(c)}$) and PLDCVMs ($D^{*+}_{(s)}, B^{*+}_{(c)}$) from Eq. (12) and (13), respectively. Thereafter, the BFs can be calculated using the observed lifetimes $(\tau_P)$ of the pseudoscalar mesons $(D_{(s)}^+, B_{(c)})$  and the total decay width $\Gamma^{total}_{D_d^{*+}}$ from Table III. In the absence of observed data on the total decay widths for the decaying vector meson, $D_s^{*+}$, we take $\Gamma^{total}_{D_s^{*+}}=0.07 \pm 0.028$ keV from the LQCD \cite{donald2014prediction} calculation. However, for $B_u^*$ and $B_c^*$, the isospin violating decay modes $B_u^*\to B_u\pi$ and $B_c^*\to B_c\pi$ are explicitly forbidden as $m_{B_u^*}-m_{B_u}\simeq45 \ MeV < m_\pi$ and $m_{B_c^*}-m_{B_c}\simeq 16 \ MeV < m_\pi$. In this sector the electromagnetic radiative transitions: $B_u^*\to B_u\gamma$ and $B_c^*\to B_c\gamma$ should be dominant decay modes. Therefore, in the present calculation, we take the assumption that $\Gamma_{B_u^*}^{total} \simeq \Gamma (B_u^*\to B_u\gamma) = 372 \pm 56$ eV and $\Gamma_{B_c^*}^{total} \simeq \Gamma (B_c^*\to B_c\gamma) = 33 \pm 5$ eV from the covariant confined quark model (CCQM) \cite{ivanov2023weak}.
 \begin{table}[!hbt]
 	\renewcommand{\arraystretch}{0.7}
 	\centering
 	\setlength\tabcolsep{10pt}
 	\caption{Ratio of decay constants for  $(B_u, B_c, B_u^*, B_c^*)$ mesons compared with the available experimental data and other theoretical predictions.}
 	\label{tab7}
 
 	\begin{tabular}{l|l|l|l}
 		\hline
 		\hline  Reference&$f_{B_u^{*+}}/f_{B_u^{+}}$& Reference& $f_{B_c^{*+}}/f_{B_c^{+}}$ \\
\hline
This work&$1.069_{+0.021-0.015}^{-0.018+0.014}$&This work&$1.037_{+0.006-0.002}^{-0.005+0.002}$\\
LFQM Lin \cite{choi2007} &1.09&LFQM Lin \cite{choi2015variational}&$1.005^{-0.052}_{+0.018}$ \\
LFQM HO \cite{choi2007}&1.07&LCDA \cite{hwang2010analyses}&1.08\\
LFQM Lin \cite{choi2015variational}&1.04&QCD SR \cite{aliev2019properties}&$1.11^{+0.1}_{-0.015}$\\
LFQM \cite{dhiman2019twist}&$1.06_{-0.013+0.011}^{+0.010-0.011}$&QCD SR \cite{narison2020spectra}&$1.19^{+0.061}_{-0.067}$\\
LCQM \cite{dhiman2018}&1.09&&\\
QCD SR \cite{gelhausen2013decay}&$1.02^{+0.02}_{-0.09}$&&\\
RQM \cite{ebert2006relativistic}&1.16&&\\
RQM \cite{hwang1997decay}&$1.09\pm0.01$&&\\
NRQM \cite{yazarloo2017mass}&0.99&&\\
NRQM \cite{abu2018heavy}&1.33&&\\
\hline
 		\hline
 	\end{tabular}
  
 \end{table}

 \begin{table}[!hbt]
	\renewcommand{\arraystretch}{1.2}
	\centering
	\setlength\tabcolsep{2pt}
	\caption{The predicted BFs ${\cal B}(P\to l^+\nu_l)$ in comparison with the observed value and other theoretical predictions.}
	\label{tab4}
	\resizebox{\textwidth}{!}
	{
		\begin{tabular}{lllllll}
			\hline
   \hline
		\ \ ${\cal B}(P\to l^+\nu_l)$&  \ \ This work&\ \ \ \ \ \cite{yazarloo2016study}& \ \ \ \ \ \ \ \ \ \ \ \cite{abu2018heavy}& \ \ \ \ \ \cite{hassanabadi2016study}&  \ \ \ \ \ \ \ \ \ \ \cite{zhou2017leptonic}& \ \ \ \ \  Expt. \cite{particle2022review}\\
			\hline
			${\cal B}(D^{+}\to e^+\nu_e)$&\ $(10.109^{+1.134+0.944}_{-1.129-0.942})\times 10^{-9}$&\  $17.7\times10^{-9}$& \ \ \ \ \ $9.84\times10^{-9}$&\ \ $11.3\times10^{-9}$&$(8.6\pm 0.5)\times 10^{-9}$&\ \ \ $<8.8\times 10^{-6}$\\
			${\cal B}(D^{+}\to \mu^+\nu_\mu)$&\ $(4.295^{+0.482+0.401}_{-0.479-0.400})\times 10^{-4}$&\  $7.54\times10^{-4}$&\ \ \ \ \ $4.29\times10^{-4}$&\ \ $4.77\times10^{-4}$&$(3.6\pm 0.2)\times 10^{-4}$& \ \ \ $(3.74\pm 0.17)\times 10^{-4}$\\
			${\cal B}(D^{+}\to \tau^+\nu_\tau)$&\ $(11.419^{+1.303+1.136}_{-1.333-1.122})\times 10^{-4}$&\  $17.9\times10^{-4}$&\ \ \ \ \ $10.55\times10^{-4}$&\ \ $20.3\times10^{-4}$&$(9.6\pm 0.6)\times 10^{-4}$& \ \ \ $(12\pm 2.7)\times 10^{-4}$\\
			\hline
   \ \ ${\cal B}(P\to l^+\nu_l)$&  \ \ This work& \ \ \ \ \cite{yazarloo2016study}& \ \ \ \ \ \ \ \ \ \ \ \cite{abu2018heavy}&  \ \ \ \ \ \cite{hassanabadi2016study}&  \ \ \ \ \ \ \ \ \ \ \cite{zhou2017leptonic}&\ \ \ \ \  Expt. \cite{particle2022review}\\
			\hline
			${\cal B}(D_s^{+}\to e^+\nu_e)$&\ $(1.298^{+0.151+0.111}_{-0.150-0.112})\times 10^{-7}$&\ $1.82\times10^{-7}$&\ \ \ \ \ \ $1.163\times10^{-7}$&\ \ $1.63\times10^{-7}$&$(1.3\pm 0.1)\times 10^{-7}$&\ \ \ $<8.3\times 10^{-5}$\\
			${\cal B}(D_s^{+}\to \mu^+\nu_\mu)$&\ $(5.517^{+0.641+0.473}_{-0.638-0.475})\times 10^{-3}$&\   $7.74\times10^{-3}$&\ \ \ \ \ \ $5.078\times10^{-3}$&\ \ $6.9\times10^{-3}$&$(5.5\pm 0.5)\times 10^{-3}$&\ \ \ $(5.43\pm 0.15)\times 10^{-3}$\\
			${\cal B}(D_s^{+}\to \tau^+\nu_\tau)$&\ $(5.408^{+0.697+0.531}_{-0.679-0.522})\times 10^{-2}$&\  $8.2\times10^{-2}$&\ \ \ \ \ \ $0.4451\times10^{-3}$&\ \ $6.49\times10^{-2}$&$(5.4\pm 0.5)\times 10^{-2}$&\ \ \ $(5.32\pm 0.11)\times 10^{-2}$\\
			\hline
   \ \ ${\cal B}(P\to l^+\nu_l)$&  \ \ This work& \ \ \ \ \ \cite{abu2018heavy}&  \ \ \ \ \ \cite{hassanabadi2016study}&  \ \ \ \ \ \ \ \ \ \ \cite{zhou2017leptonic}&\ \ \ \ \cite{kher2017spectroscopy}&\ \ \ \ \  Expt. \cite{particle2022review}\\
			\hline
			${\cal B}(B_u^{+}\to e^+\nu_e)$&\ $(6.582^{+1.074+0.928}_{-0.986-0.866})\times 10^{-12}$&\ $6.162\times10^{-12}$&\ \ $6.22\times10^{-12}$&$(8.4\pm 0.4)\times 10^{-12}$&\ \ $8.64\times10^{-12}$&\ \ $<9.8\times 10^{-7}$\\
			${\cal B}(B_u^{+}\to \mu^+\nu_\mu)$&\ $(2.812^{+0.459+0.396}_{-0.421-0.370})\times 10^{-7}$&\ $2.705\times10^{-7}$&\ \ $2.63\times10^{-7}$&$(3.5\pm 0.3)\times 10^{-7}$&\ \ $0.37\times10^{-7}$&\ \ $<8.6\times 10^{-7}$\\
			${\cal B}(B_u^{+}\to \tau^+\nu_\tau)$&\ $(6.257^{+1.022+0.883}_{-0.938-0.824})\times 10^{-5}$&\ $6.088\times10^{-5}$&\ \ $5.9\times10^{-5}$&$(8.0\pm 0.4)\times 10^{-5}$&\ \ $8.2\times10^{-5}$&\ \ $(10.9\pm 2.4)\times 10^{-5}$\\
			\hline
   \ \ ${\cal B}(P\to l^+\nu_l)$&  \ \ This work&  \ \ \ \ \ \ \ \ \ \ \cite{zhou2017leptonic}& \ \ \ \ \ \ \ \ \ \ \ \cite{sun2019}&&\ \ \ \ \ &\\
			\hline
			${\cal B}(B_c^{+}\to e^+\nu_e)$&\ $(0.748^{+0.114+0.105}_{-0.103-0.094})\times 10^{-9}$&\ \ $(2.2\pm 0.2)\times 10^{-9}$&\ \ $(2.24\pm0.24)\times 10^{-9}$&&\\
			${\cal B}(B_c^{+}\to \mu^+\nu_\mu)$&\ $(3.197^{+0.486+0.447}_{-0.439-0.401})\times 10^{-5}$&\ \ $(9.2\pm 0.9)\times 10^{-5}$&\ \ $(9.6\pm1.0)\times 10^{-5}$&&\\
			${\cal B}(B_c^{+}\to \tau^+\nu_\tau)$&\ $(0.765^{+0.116+0.107}_{-0.105-0.096})\times 10^{-2}$&\ \ $(2.2\pm 0.2)\times 10^{-2}$&\ \ $(2.29\pm0.24)\times 10^{-2}$&&\\
   
			\hline
			\hline
   
		\end{tabular}
	}

\end{table}

	\begin{table}[!hbt]
	\renewcommand{\arraystretch}{1.25}
	\centering
	\setlength\tabcolsep{2pt}
	\caption{The predicted BFs ${\cal B}(V\to l^+\nu_l)$ in comparison with the observed value and other theoretical predictions.}
	\label{tab5}
	\resizebox{\textwidth}{!}
	{
		\begin{tabular}{lllll}
			\hline
   \hline
			${\cal B}(V\to l^+\nu_l)$& \ \  This work& \ \ \ \ \ \ \ \ \ \ \cite{yang2021purely}& \ \ \ \ \ \ \ \ \ \  \cite{zhou2017leptonic}&  \\
			\hline
			${\cal B}(D^{*+}\to e^+\nu_e)$&\ \ $(11.655^{+0.505+0.966}_{-0.658-1.117})\times 10^{-10}$&$ ({9.5}^{+2.9}_{-2.4})\times10^{-10}$&$(11\pm 1)\times 10^{-10}$&\\
			${\cal B}(D^{*+}\to \mu^+\nu_\mu)$&\ \ $(11.607^{+0.503+0.963}_{-0.655-1.113})\times 10^{-10}$&$ ({9.5}^{+2.9}_{-2.4})\times10^{-10}$&$(11\pm 1)\times 10^{-10}$&\\
			${\cal B}(D^{*+}\to \tau^+\nu_\tau)$&\ \ $(0.775^{+0.334+0.641}_{-0.430-0.736})\times 10^{-10}$&$ (0.6 \pm 0.2)\times10^{-10}$&$(0.72\pm 0.08)\times 10^{-10}$&\\
			\hline
   ${\cal B}(V\to l^+\nu_l)$& \ \  This work& \ \ \ \ \ \ \ \ \ \ \cite{yang2021purely}& \ \ \ \ \ \ \ \ \ \  \cite{zhou2017leptonic}& \ \ \ \ \ \ \ \ \ \ \ \ \ Expt. \cite{ablikim2023first}\\
   \hline
			${\cal B}(D_s^{*+}\to e^+\nu_e)$&\ \ $(3.765^{-0.883-0.828}_{+2.057+1.932})\times 10^{-5}$&$(6.7\pm 0.4)\times10^{-6}$&$(3.1\pm 0.4)\times 10^{-6}$&$(2.1^{+1.2}_{{-0.9}_{stat.}}\pm {0.2}_{syst.})\times 10^{-5}$\\
			${\cal B}(D_s^{*+}\to \mu^+\nu_\mu)$&\ \ $(3.751^{-0.880-0.825}_{+2.049+1.925})\times 10^{-5}$&$(6.7\pm 0.4)\times10^{-6}$&$(3.1\pm 0.4)\times 10^{-6}$&\\
			${\cal B}(D_s^{*+}\to \tau^+\nu_\tau)$&\ \ $(0.436^{-0.102-0.095}_{+0.237+0.223})\times 10^{-5}$&$(0.78\pm 0.04)\times10^{-6}$&$(0.36\pm 0.04)\times 10^{-6}$&\\
			\hline
   ${\cal B}(V\to l^+\nu_l)$& \ \  This work& \ \ \ \ \ \ \ \ \ \ \cite{yang2021purely}& \ \ \ \ \ \ \ \ \ \  \cite{zhou2017leptonic}& \ \ \ \ \ \ \ \ \ \ \cite{sun2019}  \\
   \hline
			${\cal B}(B_u^{*+}\to e^+\nu_e)$&\ \ $(5.942^{-0.149+0.091}_{+0.261-0.022})\times 10^{-10}$&$(3.0\pm 0.4)\times10^{-10}$&$(6.4\pm 2.6)\times 10^{-11}$&$(9.0\pm 2.5)\times 10^{-10}$\\
			${\cal B}(B_u^{*+}\to \mu^+\nu_\mu)$&\ \ $(5.938^{-0.149+0.091}_{+0.261-0.023})\times 10^{-10}$&$(3.0\pm 0.4)\times10^{-10}$&$(6.4\pm 2.6)\times 10^{-11}$&$(9.0\pm 2.5)\times 10^{-10}$\\
			${\cal B}(B_u^{*+}\to \tau^+\nu_\tau)$&\ \ $(4.953^{-0.124+0.076}_{+0.217-0.019})\times 10^{-10}$&$(2.5\pm 0.4)\times10^{-10}$&$(5.4\pm 2.2)\times 10^{-11}$&$(7.5\pm 2.1)\times 10^{-10}$\\
			\hline
   ${\cal B}(V\to l^+\nu_l)$& \ \  This work& \ \ \ \ \ \ \ \ \ \ \cite{yang2021purely}& \ \ \ \ \ \ \ \ \ \  \cite{zhou2017leptonic}&   \\
   \hline
			${\cal B}(B_c^{*+}\to e^+\nu_e)$&\ \ $(3.186^{-0.082-0.076}_{+0.150+0.144})\times 10^{-6}$& \ \ $3.8^{+0.4}_{-0.3}\times10^{-6}$&$(4.3\pm 0.4)\times 10^{-6}$&\\
			${\cal B}(B_c^{*+}\to \mu^+\nu_\mu)$&\ \ $(3.184^{-0.082-0.076}_{+0.150+0.144})\times 10^{-6}$&\ \ $3.8^{+0.4}_{-0.3}\times10^{-6}$&$(4.3\pm 0.4)\times 10^{-6}$&\\
			${\cal B}(B_c^{*+}\to \tau^+\nu_\tau)$&\ \ $(2.813^{-0.080-0.075}_{+0.124+0.119})\times 10^{-6}$&\ \ $3.3^{+0.4}_{-0.3}\times10^{-6}$&$(3.8\pm 0.4)\times 10^{-6}$&\\
			\hline
		\hline	
		\end{tabular}
	}
\end{table}
\par Our predicted BFs for the PLDCPMs and PLDCVMs are listed in Table VIII and IX, respectively in comparison with the available observed data and other model predictions. While calculating the uncertainties in our predictions for BFs, we take into account the uncertainties of the relevant physical quantities such as decaying meson masses $m_{P(V)}$, lepton masses $m_l$ $(l=e, \mu, \tau)$, lifetimes $\tau_P$, total decay widths $\Gamma_V^{total}$, CKM parameters $|V_{q_1q_2}|$ and our predicted decay constants $f_{P(V)}$. 
\par As shown in Table VIII, our predicted BFs for PLDCPMs: ${\cal B}(D_{(s)}^{+}\to l^+\nu_l)$, ${\cal B}(B_u^{+}\to e^+\nu_e)$ and ${\cal B}(B_u^{+}\to \mu^+\nu_\mu)$ are obtained well within the experimental limit and in reasonable agreement with other theoretical results of Ref. \cite{abu2018heavy, hassanabadi2016study, zhou2017leptonic}. Our predicted ${\cal B}(B_u^{+}\to \tau^+\nu_\tau)$ is also comparable to the observed data within the experimental uncertainties and the theoretical predictions of \cite{abu2018heavy, hassanabadi2016study}. We predict the BFs for $B_c$ meson decay in the same order of magnitude, although in terms of their absolute values, our results are underestimated as compared to the available theoretical results of \cite{zhou2017leptonic, sun2019}. As one can see, our predicted  ${\cal B}(D_s^{*+}\to e^+\nu_e)$ agree with the recently observed data from BESIII Collaboration within the experimental uncertainties and compares well with the prediction of Ref. \cite{zhou2017leptonic}. In the PLDCVMs, our predictions for ${\cal B}(D^{*+}\to l^+\nu_l)$, ${\cal B}(B_u^{*+}\to l^+\nu_l)$, ${\cal B}(B_c^{*+}\to l^+\nu_l)$, $(l=e, \mu, \tau)$ and ${\cal B}(D_s^{*+}\to l^+\nu_l)$, ($l=\mu, \tau$), are in good agreement with those of Ref. \cite{yang2021purely, zhou2017leptonic}, Ref. \cite{zhou2017leptonic}, Ref. \cite{yang2021purely} and Ref. \cite{zhou2017leptonic}, respectively. Finally, we calculate the ratios of BFs: $({\cal R}_\mu^{\tau})^D$ and $({\cal R}_\mu^{\tau})^{D_s}$. Our predictions of $({\cal R}_\mu^{\tau})^D=2.66$ and $({\cal R}_\mu^{\tau})^{D_s}=9.80$ are consistent with the corresponding observed data $3.21\pm 0.73$ and $9.82\pm 0.40$, respectively from PDG \cite{particle2022review}.
\section{Summary and Conlusion}
In the present work, we study the purely leptonic decays of heavy-flavored charged pseudoscalar and vector ($D_{(s)}^{(*)+}, B_{(c)}^{(*)+}$) mesons in the framework of the relativistic independent quark (RIQ) model based on an average flavor-independent confining potential in the scalar-vector harmonic form. Using the meson wave functions derivable in the RIQ model, we first compute the mass spectra in reasonable agreement with the observed data for the masses of the ground state pseudoscalar ($J^P=0^-$) and vector ($J^P=1^-$) mesons. With the model parameters: quark masses $m_q$ and potential parameters ($V_0, a$), as fixed from the hadron spectroscopy, we predict the decay constants: $f_P$ and $f_V$, for purely leptonic decays of charged pseudoscalar (P) and vector (V) mesons, respectively.
\par To study the sensitivity of the input parameters in our predictions of decay constants $f_P$ and $f_V$, we include the systematic errors in our analysis both from the $\pm10\%$ variation of the potential parameters ($V_0, a$) for fixed quark masses $m_q$ and $\pm10\%$ variation of quark masses $m_q$ for fixed values of potential parameters ($V_0, a$). We find that our predicted decay constants $f_V$s for the PLDCVMs are more sensitive to the variation of potential parameters with fixed quark masses. However, our predicted $f_P$s are found more sensitive to the variation of quark mass value with fixed potential parameters.
\par Our predicted decay constants $f_{D^+}$, $f_{D_s^+}$ and $f_{D_s^{*+}}$ not only agree with the corresponding observed data within their experimental limits but compare well with several SM predictions. Our result for $f_D^{*+}$ is also in good comparison with the results of LFQM Lin., LFQM, QCD SR and LQCD calculations. Our prediction of $f_{B_u^+}$ lies well within the experimental limit. In the present study, our predicted decay constants $f_{B_u^+}$ and $f_{B_u^{*+}}$ compare well with the predictions of BS, LFQM, LQCD and those of LFQM and LQCD, respectively. In the absence of the observed data in $B_c^{*+}$ sector, we compare our predicted decay constant with a few theoretical results available in the literature. Our result for $f_{B_c^+}$ is comparable to that of QCD SR and that of $f_{B_c^{*+}}$ appears somewhat underestimated compared to the results of BS, LFQM Lin, LCDA and QCD SR.
\par We also calculate the ratios: $f_V/f_P$, $f_{P_1}/f_{P_2}$, $f_{V_1}/f_{V_2}$, which are sensitive to the difference between the vector (V) and pseudoscalar (P), pseudoscalar ($P_1$) and pseudoscalar ($P_2$), vector ($V_1$) and vector ($V_2$) wave function, respectively. While our predicted $f_{D_s^{*+}}/f_{D_s^{+}}$ matches exactly with the LFQM Lin. prediction, our results for $f_{D_s^{*+}}/f_{D_s^{+}}$ and $f_{D^{*+}}/f_{D^{+}}$ agree with those of theoretical predictions based on RQM, LFQM, QCD SR and LQCD calculations. We find $f_{D_s^{+}}/f_{D^{+}}$ in good comparison with the observed data. Our predicted $f_{D_s^{+}}/f_{D^{+}}$ is also consistent with other SM predictions including those obtained from the RQM, LFHQCD and QCD SR approaches and our result for $f_{D_s^{*+}}/f_{D^{*+}}$ compares well with the finding of LFQM Lin. and LFQM calculations. In the $b$-flavored meson sector, our results for $f_{B_u^{*+}}/f_{B_u^{+}}$ is comparable to the predictions of LFQM Lin. LFQM  and that for $f_{B_c^{*+}}/f_{B_c^{+}}$ is in reasonable agreement with the results of LFQM Lin. and QCD SR calculation. Our predictions on $f_{B_c^{*+}}/f_{B_u^{*+}}$ and $f_{B_c^{+}}/f_{B_u^{+}}$, can be verified in the upcoming experimental measurement and compared with future SM predictions.  Finally, we predict the BFs of PLDCMs. Our predicted BFs for PLDCPMs: ${\cal B}(D_{(s)}^{+}\to l^+\nu_l)$, ${\cal B}(B_{(c)}^{+}\to l^+\nu_l), l=e, \mu, \tau$ are obtained within experimental limits and also in agreement with theoretical results based on the NRQM, RQM and LFQM calculations. In the PLDCVMs, our prediction of ${\cal B}(D_{s}^{*+}\to e^+\nu_e)$ agrees with the recently observed data from BESIII Collaboration. For other modes such as $D_{(s)}^{*+}\to l^+\nu_l$, $B_{(c)}^{*+}\to l^+\nu_l$, our predicted BFs in this sector find good agreement with those of LFQM and LQCD calculation. 
\par With the potential prospects of precision measurements, high data statistics and improved analytical tools available in the high-energy experimental frontiers, careful measurement of yet unmeasured decay constants $f_{P(V)}$ and BFs might lead to predictions even close to the accessible limits of ongoing experiments and their upgrades at Belle-II, SCTF or STCF, CEPC, FCC-ee and LHCb. In heavy-flavor sector, the PLDCMs thus would continue to provide a fascinating area of experimental and theoretical research in the future.

\section*{Acknowledgments}
The library and computational facilities provided by authorities of Siksha 'O' Anusandhan Deemed to be University, Bhubaneswar, 751030, India are duly acknowledged.

\bibliography{ref}{}

\end{document}